\pdfoutput=1
\documentclass[12pt,a4paper]{article}
\usepackage{ifthen} % for conditional statements
\newboolean{pdflatex}
\setboolean{pdflatex}{true} % False for eps figures

\newboolean{articletitles}
\setboolean{articletitles}{true} % False removes titles in references

\newboolean{uprightparticles}
\setboolean{uprightparticles}{false} %True for upright particle symbols
\usepackage{afterpage}

\def\paperauthors{LHCb collaboration} % Leave as is for PAPER, CONF and FIGURE
\def\paperasciititle{First branching fraction measurement of the suppressed decay Xic0->pi-Lc+
} % Set ASCII title hereLHCb-PAPER-2015-041
\def\papertitle{First branching fraction measurement of the suppressed decay $\decXicz$
} % Latex formatted title
\def\paperkeywords{{High Energy Physics}, {LHCb}} % Comma separated list
\def\papercopyright{\the\year\ CERN for the benefit of the LHCb collaboration} % new since 9/Apr/2018
\def\paperlicence{CC BY 4.0 license}
\def\paperlicenceurl{https://creativecommons.org/licenses/by/4.0/}
\usepackage[top=1in, bottom=1.25in, left=1in, right=1in]{geometry}

\usepackage{microtype}
\usepackage{lineno}  % for line numbering during review
\usepackage{xspace} % To avoid problems with missing or double spaces after
\usepackage{caption} %these three command get the figure and table captions automatically small

\usepackage{graphicx}  % to include figures (can also use other packages)
\usepackage{color}
\usepackage{colortbl}
\graphicspath{{./figs/}} % Make Latex search fig subdir for figures
\DeclareGraphicsExtensions{.pdf,.PDF,png,.PNG}

\usepackage{amsmath} % Adds a large collection of math symbols
\usepackage{amssymb}
\usepackage{amsfonts}
\usepackage{upgreek} % Adds in support for greek letters in roman typeset

\newcommand*\patchAmsMathEnvironmentForLineno[1]{%
\expandafter\let\csname old#1\expandafter\endcsname\csname #1\endcsname
\expandafter\let\csname oldend#1\expandafter\endcsname\csname
end#1\endcsname
 \renewenvironment{#1}%
   {\linenomath\csname old#1\endcsname}%
   {\csname oldend#1\endcsname\endlinenomath}%
}
\newcommand*\patchBothAmsMathEnvironmentsForLineno[1]{%
  \patchAmsMathEnvironmentForLineno{#1}%
  \patchAmsMathEnvironmentForLineno{#1*}%
}
\AtBeginDocument{%
\patchBothAmsMathEnvironmentsForLineno{equation}%
\patchBothAmsMathEnvironmentsForLineno{align}%
\patchBothAmsMathEnvironmentsForLineno{flalign}%
\patchBothAmsMathEnvironmentsForLineno{alignat}%
\patchBothAmsMathEnvironmentsForLineno{gather}%
\patchBothAmsMathEnvironmentsForLineno{multline}%
\patchBothAmsMathEnvironmentsForLineno{eqnarray}%
}

\usepackage{hyperxmp}

\usepackage[pdftex,
            pdfauthor={\paperauthors},
            pdftitle={\paperasciititle},
            pdfkeywords={\paperkeywords},
            pdfcopyright={Copyright (C) \papercopyright},
            pdflicenseurl={\paperlicenceurl}]{hyperref}

\usepackage[all]{hypcap} % Internal hyperlinks to floats.

\usepackage{xspace} 
\usepackage{upgreek}

\def\lhcb   {\mbox{LHCb}\xspace}

\def\MagUp {\mbox{\em Mag\kern -0.05em Up}\xspace}

\ifthenelse{\boolean{uprightparticles}}%
{

 \def\Ppi         {\ensuremath{\uppi}\xspace}

 \def\PDelta      {\ensuremath{\Delta}\xspace}                 
 \def\PXi         {\ensuremath{\Xi}\xspace}                 
 \def\PLambda     {\ensuremath{\Lambda}\xspace}                 
 \def\PSigma      {\ensuremath{\Sigma}\xspace}                 
 \def\POmega      {\ensuremath{\Omega}\xspace}                 
 \def\PUpsilon    {\ensuremath{\Upsilon}\xspace}

 \def\PB      {\ensuremath{\mathrm{B}}\xspace}                 
                  
 \def\PD      {\ensuremath{\mathrm{D}}\xspace}

 \def\PK      {\ensuremath{\mathrm{K}}\xspace}

 \def\Pb      {\ensuremath{\mathrm{b}}\xspace}                 
 \def\Pc      {\ensuremath{\mathrm{c}}\xspace}                 
 \def\Pd      {\ensuremath{\mathrm{d}}\xspace}

 \def\Pi      {\ensuremath{\mathrm{i}}\xspace}

 \def\Pp      {\ensuremath{\mathrm{p}}\xspace}

 \def\Ps      {\ensuremath{\mathrm{s}}\xspace}                 
                  
 \def\Pu      {\ensuremath{\mathrm{u}}\xspace}

 \def\thebaroffset{0.0em}
}
{

 \def\Ppi         {\ensuremath{\pi}\xspace}

 \mathchardef\PDelta="7101
 \mathchardef\PXi="7104
 \mathchardef\PLambda="7103
 \mathchardef\PSigma="7106
 \mathchardef\POmega="710A
 \mathchardef\PUpsilon="7107
                  
 \def\PB      {\ensuremath{B}\xspace}                 
                  
 \def\PD      {\ensuremath{D}\xspace}

 \def\PK      {\ensuremath{K}\xspace}

 \def\Pb      {\ensuremath{b}\xspace}                 
 \def\Pc      {\ensuremath{c}\xspace}                 
 \def\Pd      {\ensuremath{d}\xspace}

 \def\Pi      {\ensuremath{i}\xspace}

 \def\Pp      {\ensuremath{p}\xspace}

 \def\Ps      {\ensuremath{s}\xspace}                 
                  
 \def\Pu      {\ensuremath{u}\xspace}

}

\makeatletter
\ifcase \@ptsize \relax% 10pt
  \newcommand{\miniscule}{\@setfontsize\miniscule{4}{5}}% \tiny: 5/6
\or% 11pt
  \newcommand{\miniscule}{\@setfontsize\miniscule{5}{6}}% \tiny: 6/7
\or% 12pt
  \newcommand{\miniscule}{\@setfontsize\miniscule{5}{6}}% \tiny: 6/7
\fi
\makeatother

\DeclareRobustCommand{\optbar}[1]{\shortstack{{\miniscule (\rule[.5ex]{1.25em}{.18mm})}
  \\ [-.7ex] $#1$}}

\def\uquark    {{\ensuremath{\Pu}}\xspace}
\def\uquarkbar {{\ensuremath{\overline \uquark}}\xspace}

\def\dquark    {{\ensuremath{\Pd}}\xspace}

\def\squark    {{\ensuremath{\Ps}}\xspace}

\def\cquark    {{\ensuremath{\Pc}}\xspace}

\def\bquark    {{\ensuremath{\Pb}}\xspace}

\def\pion   {{\ensuremath{\Ppi}}\xspace}

\def\pip    {{\ensuremath{\pion^+}}\xspace}
\def\pim    {{\ensuremath{\pion^-}}\xspace}

\def\kaon    {{\ensuremath{\PK}}\xspace}
\def\KorKbar {\kern \thebaroffset\optbar{\kern -\thebaroffset \PK}{}\xspace}

\def\Km      {{\ensuremath{\kaon^-}}\xspace}

\def\D       {{\ensuremath{\PD}}\xspace}

\def\DorDbar {\kern \thebaroffset\optbar{\kern -\thebaroffset \PD}\xspace}
\def\Dz      {{\ensuremath{\D^0}}\xspace}

\def\Dp      {{\ensuremath{\D^+}}\xspace}
\def\Dm      {{\ensuremath{\D^-}}\xspace}

\def\DpDm    {\ensuremath{\Dp {\kern -0.16em \Dm}}\xspace}

\def\Ds      {{\ensuremath{\D^+_\squark}}\xspace}

\def\B       {{\ensuremath{\PB}}\xspace}

\def\BorBbar {\kern \thebaroffset\optbar{\kern -\thebaroffset \PB}\xspace}
\def\Bz      {{\ensuremath{\B^0}}\xspace}

\def\Bd      {{\ensuremath{\B^0}}\xspace}

\def\BdorBdbar {\kern \thebaroffset\optbar{\kern -\thebaroffset \Bd}\xspace}
\def\Bu      {{\ensuremath{\B^+}}\xspace}

\def\Bp      {{\ensuremath{\Bu}}\xspace}

\def\Bd      {{\ensuremath{\B^0}}\xspace}
\def\Bs      {{\ensuremath{\B^0_\squark}}\xspace}

\def\BsorBsbar {\kern \thebaroffset\optbar{\kern -\thebaroffset \Bs}\xspace}

  \def\Y#1S{\ensuremath{\PUpsilon{(#1S)}}\xspace}% no space before {...}!

\def\proton      {{\ensuremath{\Pp}}\xspace}

\def\Xires       {{\ensuremath{\PXi}}\xspace}

\def\Lz          {{\ensuremath{\PLambda}}\xspace}
\def\Lbar        {{\ensuremath{\kern 0.1em\overline{\kern -0.1em\PLambda}}}\xspace}
\def\LorLbar    {\kern 0.18em\optbar{\kern -0.18em \PLambda}{}\xspace}

\def\Lb      {{\ensuremath{\Lz^0_\bquark}}\xspace}

\def\Lc      {{\ensuremath{\Lz^+_\cquark}}\xspace}

\def\Xib     {{\ensuremath{\Xires_\bquark}}\xspace}
\def\Xibz    {{\ensuremath{\Xires^0_\bquark}}\xspace}
\def\Xibm    {{\ensuremath{\Xires^-_\bquark}}\xspace}
\def\Xibprm  {{\ensuremath{\Xires^{\prime -}_\bquark}}\xspace}  % for this paper
\def\Xibstm  {{\ensuremath{\Xires^{* -}_\bquark}}\xspace}  % for this paper

\def\Xicz    {{\ensuremath{\Xires^0_\cquark}}\xspace}
\def\Xicp    {{\ensuremath{\Xires^+_\cquark}}\xspace}

\def\Lb           {{\ensuremath{\Lz^0_\bquark}}\xspace}

\def\Xib          {{\ensuremath{\Xires_\bquark}}\xspace}
\def\Xibz         {{\ensuremath{\Xires^0_\bquark}}\xspace}
\def\Xibm         {{\ensuremath{\Xires^-_\bquark}}\xspace}

\def\BF         {{\ensuremath{\mathcal{B}}}\xspace}

\newcommand{\decay}[2]{\ensuremath{#1\!\to #2}\xspace} 

\def\to                 {\ensuremath{\rightarrow}\xspace}

\def\AT#1     {\ensuremath{A_{\mathrm{T}}^{#1}}\xspace}           % 2

\def\decXicz       {\decay{\Xicz}{\pim \Lc}}

\def\decXicp       {\decay{\Xicp}{\proton \Km \pip}}
\def\decLc         {\decay{\Lc}{\proton \Km \pip}}

\def\C#1      {\ensuremath{\mathcal{C}_{#1}}\xspace}                       % 9
\def\Cp#1     {\ensuremath{\mathcal{C}_{#1}^{'}}\xspace}                    % 7
\def\Ceff#1   {\ensuremath{\mathcal{C}_{#1}^{\mathrm{(eff)}}}\xspace}        % 9
\def\Cpeff#1  {\ensuremath{\mathcal{C}_{#1}^{'\mathrm{(eff)}}}\xspace}       % 7
\def\Ope#1    {\ensuremath{\mathcal{O}_{#1}}\xspace}                       % 2
\def\Opep#1   {\ensuremath{\mathcal{O}_{#1}^{'}}\xspace}                    % 7

\newcommand{\aunit}[1]{\ensuremath{\text{\,#1}}}       
\newcommand{\tev}{\aunit{Te\kern -0.1em V}\xspace}
\newcommand{\gev}{\aunit{Ge\kern -0.1em V}\xspace}
\newcommand{\mev}{\aunit{Me\kern -0.1em V}\xspace}
\newcommand{\kev}{\aunit{ke\kern -0.1em V}\xspace}
\newcommand{\ev}{\aunit{e\kern -0.1em V}\xspace}
 
\newcommand{\mevc}{\ensuremath{\aunit{Me\kern -0.1em V\!/}c}\xspace}
\newcommand{\gevc}{\ensuremath{\aunit{Ge\kern -0.1em V\!/}c}\xspace}
\newcommand{\mevcc}{\ensuremath{\aunit{Me\kern -0.1em V\!/}c^2}\xspace}
\newcommand{\gevcc}{\ensuremath{\aunit{Ge\kern -0.1em V\!/}c^2}\xspace}
\def\fs   {\aunit{fs}}

\newcommand{\chisq}{\ensuremath{\chi^2}\xspace}

\def\gsim{{~\raise.15em\hbox{$>$}\kern-.85em
          \lower.35em\hbox{$\sim$}~}\xspace}
\def\lsim{{~\raise.15em\hbox{$<$}\kern-.85em
          \lower.35em\hbox{$\sim$}~}\xspace}

\def\sqs   {\ensuremath{\protect\sqrt{s}}\xspace}

\def\pt         {\ensuremath{p_{\mathrm{T}}}\xspace}

\newcommand{\lum} {\ensuremath{\mathcal{L}}\xspace}
\def\evtgen     {\mbox{\textsc{EvtGen}}\xspace}

\def\geant      {\mbox{\textsc{Geant4}}\xspace}

\def\photos     {\mbox{\textsc{Photos}}\xspace}

\def\pythia     {\mbox{\textsc{Pythia}}\xspace}

\def\tell1  {TELL1\xspace}
\def\ukl1   {UKL1\xspace}

\newcommand{\etal}{\mbox{\itshape et al.}\xspace}
\newcommand{\etc}{\mbox{\itshape etc.}\xspace}

\usepackage{cite} % Allows for ranges in citations
\usepackage{mciteplus}

\usepackage{longtable} % only for template; not usually to be used in PAPERs

\renewcommand{\thefootnote}{\fnsymbol{footnote}}
\begin{document}

% %%%%%%% CHOOSE TITLE PAGE--------
%\onecolumn
%\input{title-LHCb-INT}
%\input{title-LHCb-ANA}
%\input{title-LHCb-CONF}
%\input{title-LHCb-FIGURE}
% ===============================================================================
% Purpose: LHCb-PAPER journal paper title page template
% Author: 
% Created on: 2010-09-25
% ===============================================================================

%%%%%%%%%%%%%%%%%%%%%%%%%
%%%%%  TITLE PAGE  %%%%%%
%%%%%%%%%%%%%%%%%%%%%%%%%
\begin{titlepage}
\pagenumbering{roman}

% Header ---------------------------------------------------
\vspace*{-1.5cm}
\centerline{\large EUROPEAN ORGANIZATION FOR NUCLEAR RESEARCH (CERN)}
\vspace*{1.5cm}
\noindent
\begin{tabular*}{\linewidth}{lc@{\extracolsep{\fill}}r@{\extracolsep{0pt}}}
\ifthenelse{\boolean{pdflatex}}% Logo format choice
{\vspace*{-1.5cm}\mbox{\!\!\!\includegraphics[width=.14\textwidth]{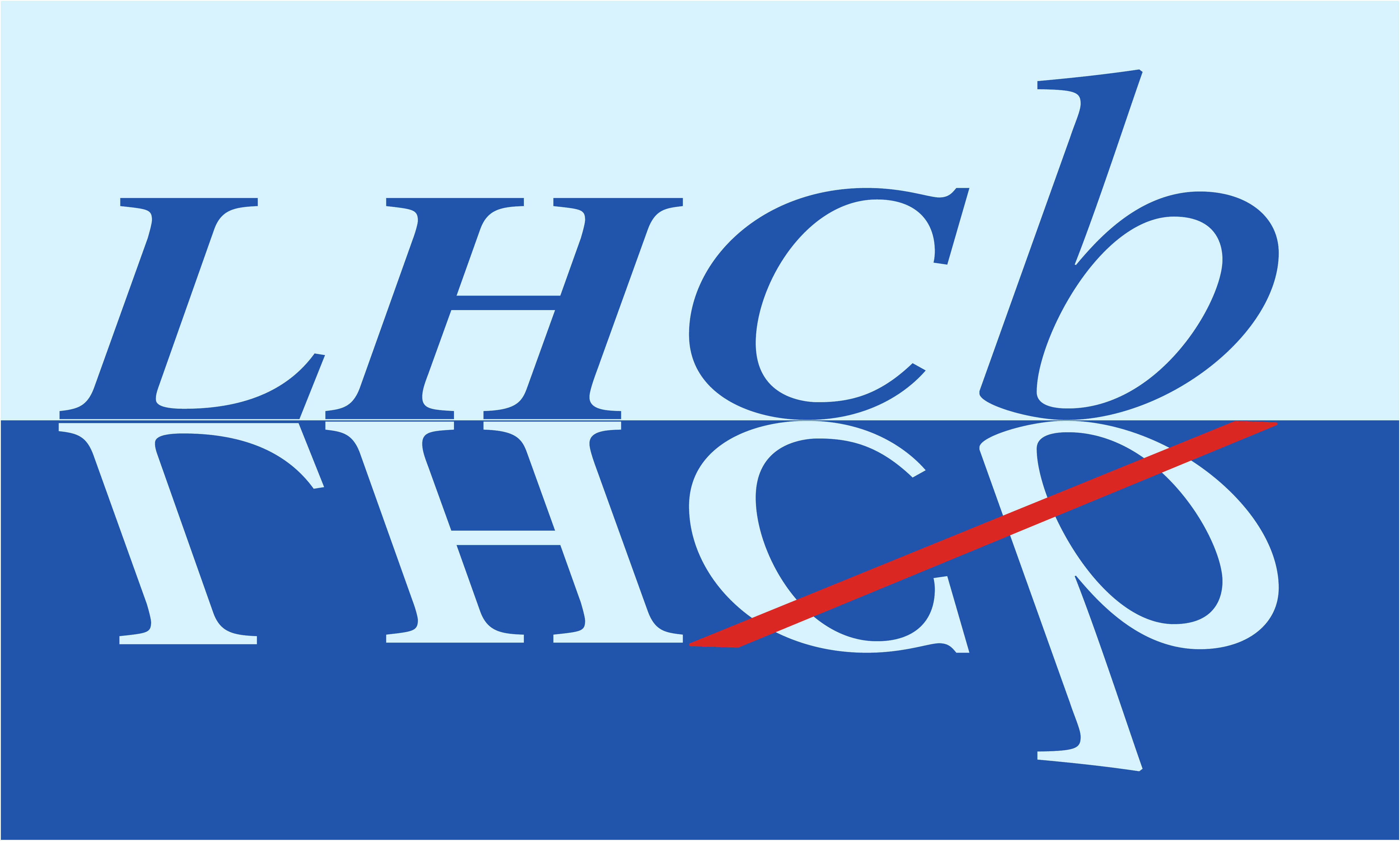}} & &}%
{\vspace*{-1.2cm}\mbox{\!\!\!\includegraphics[width=.12\textwidth]{lhcb-logo.eps}} & &}%
\\
 & & CERN-EP-2020-129 \\  % ID 
 & & LHCb-PAPER-2020-016 \\  % ID 
 & & \today \\ % Date - Can also hardwire e.g.: 23 March 2010
 & & \\
% not in paper \hline
\end{tabular*}

\vspace*{4.0cm}

% Title --------------------------------------------------
{\normalfont\bfseries\boldmath\huge
\begin{center}
% DO NOT EDIT HERE. Instead edit macro in main.tex to keep metadata correct
  \papertitle 
\end{center}
}

\vspace*{2.0cm}

% Authors -------------------------------------------------
\begin{center}
%In the footnote, replace 'paper' by 'Letter' in case of submission to PRL or PLB 
% Edit macro in main.tex to keep metadata correct
\paperauthors\footnote{Authors are listed at the end of this paper.}
\end{center}

\vspace{\fill}

% Abstract -----------------------------------------------
\begin{abstract}
  \noindent
The \Xicz baryon is unstable and usually decays into charmless final states by the $c \to s u\overline{d}$ transition. It can, however, also 
 disintegrate into a $\pi^-$ meson and a $\Lc$ baryon via $s$ quark decay or via $cs\to d c$ weak scattering. The interplay between the latter two processes governs the size of the branching fraction ${\cal{B}}(\decXicz)$, first measured here to be $(0.55\pm 0.02 \pm 0.18)$\%, where the first uncertainty is statistical and second systematic. This result is compatible with the larger  of the theoretical predictions that connect  models of hyperon decays 
 using partially conserved axial currents and SU(3) symmetry 
 with those involving the heavy-quark expansion and heavy-quark symmetry. In addition, the branching fraction of the normalization channel,
 $\BF(\decXicp) = (1.135 \pm 0.002 \pm 0.387)$\% is measured.

\end{abstract}

\vspace*{2.0cm}

\begin{center}
  Submitted to Phys.~Rev.~D Rapid Communications

\end{center}

\vspace{\fill}

{\footnotesize 
% Edit macro in main.tex to keep metadata correct
\centerline{\copyright~\papercopyright. \href{\paperlicenceurl}{\paperlicence}.}}
\vspace*{2mm}

\end{titlepage}

%%%%%%%%%%%%%%%%%%%%%%%%%%%%%%%%
%%%%%  EOD OF TITLE PAGE  %%%%%%
%%%%%%%%%%%%%%%%%%%%%%%%%%%%%%%%

%  empty page follows the title page ----
\newpage
\setcounter{page}{2}
\mbox{~}

%\twocolumn
% %%%%%%%%%%%%% ---------

\renewcommand{\thefootnote}{\arabic{footnote}}
\setcounter{footnote}{0}

%%%%%%%%%%%%%%%%%%%%%%%%%%%%%%%%
%%%%%  Table of Content   %%%%%%
%%%%%%%%%%%%%%%%%%%%%%%%%%%%%%%%
%%%% Uncomment next 2 lines if desired
%\tableofcontents
%\cleardoublepage

%%%%%%%%%%%%%%%%%%%%%%%%%
%%%%% Main text %%%%%%%%%
%%%%%%%%%%%%%%%%%%%%%%%%%

\pagestyle{plain} % restore page numbers for the main text
\newpage
\setcounter{page}{1}
\pagenumbering{arabic}

\begin{figure}[b]
\vspace{-6mm}
  \begin{center}
\includegraphics[width=0.75\linewidth]{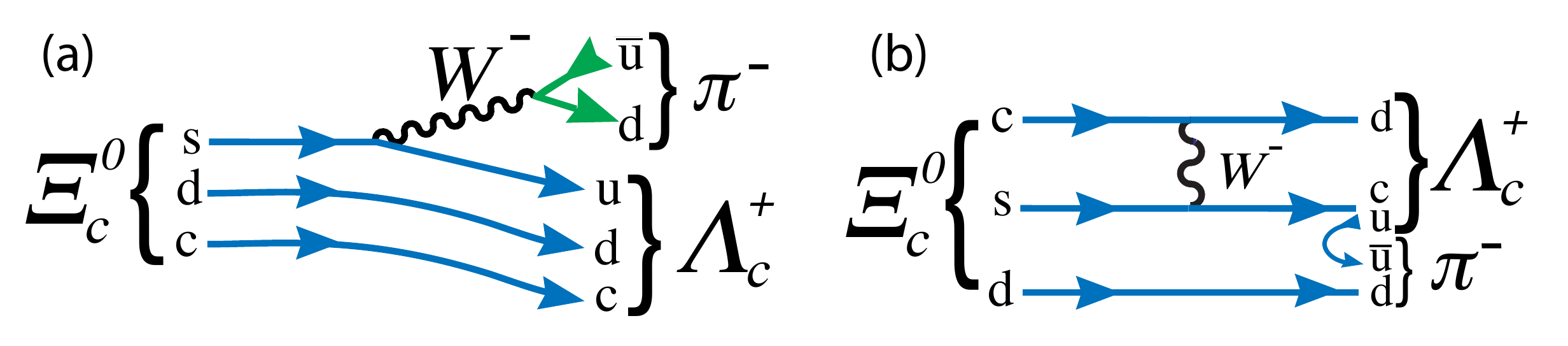}
  \end{center}
  \vspace{-6mm}
  \caption{Decay diagrams for $\Xicz\to\pi^-\PLambda_c^+$ transitions. (a) The SUUD amplitude, and (b) the WS amplitude.  }
   \label{Xic0-diag}
\end{figure}

Baryons containing both an $s$ quark and a heavy $c$ or $b$ quark, denoted as $Q$, usually decay via the disintegration of the heavy quark. There is, however, the possibility of $s$ quark decay causing the transformation.
Theoretical predictions concerning the decay widths of $\PXi_Q\to\pi\PLambda_Q$ transitions
are based on the size of the $s$ quark decay amplitude $\squark \to \uquark (\uquarkbar \dquark)$ (SUUD)
and the weak scattering  (WS) amplitude  $Q \squark \to \dquark Q$ \cite{Voloshin:2000et}.
Feynman diagrams corresponding to these amplitudes are shown in Fig.~\ref{Xic0-diag} for $\Xicz$ decay.

Studies of these $\PXi_Q$ baryon decays provide a connection to theories concerning hyperon decays
with those for the heavy \bquark and \cquark quarks.
The former use partially conserved axial currents (PCAC) and SU(3) symmetry \cite{Treiman:1967cua},
whereas the latter apply more modern approaches using four-quark operators, including the heavy quark expansion, and heavy-quark symmetry (HQS). As the $\Xibm$ baryon consists of \bquark, \squark, and \dquark quarks,
the WS amplitude is not present in $\Xibm\to\pi^-\Lb$ decays, so the measurement of that decay rate can be used to determine the SUUD amplitude.  This information can be used to predict the $\Xicz$ decay rate that, in principle, involves both amplitudes. Whenever a specific final state is mentioned additional use of the charge-conjugated state is implied.

The well-known \Xicz baryon consists of the \cquark, \squark, and \dquark quarks, and has a
lifetime of $154.5\pm 1.7 \pm 1.6\pm 1.0 \fs$ \cite{Aaij:2019lwg}.
The branching fraction ${\cal{B}}(\decXicz)$ has not been previously measured.
Several authors have made predictions using the measured SUUD amplitude
and the measured lifetimes of the SU(3) triplet baryons \Xicz, \Lc, and \Xicp, as input for determining the WS amplitude.
This method was pioneered by Voloshin~\cite{Voloshin:2000et} where he used SU(3) symmetry, PCAC and the heavy-quark limit to determine an upper limit on  $\Gamma\left(\Xibm\to\pim\Lb\right)$.
In a subsequent paper, he uses the input from the \lhcb measurement of $\BF\left(\Xibm\to\pim\Lb\right)=(0.60\pm 0.18)\%$ \cite{Aaij:2015yoy} and updated values for the charmed baryon lifetimes to find the SUUD rate and then calculates the WS amplitude.
He predicts $\BF\left(\decXicz\right) \gtrapprox (0.25\pm0.15)\cdot 10^{-3}$ \cite{Voloshin:2019ngb},
assuming negative interference between the two strangeness-changing amplitudes.

Gronau and Rosner, using the same approach as Voloshin, predict two possible branching fractions for \decXicz decay,
depending on the sign of the interference between the two decay amplitudes \cite{Gronau:2016xiq}.
Based on the measured $\BF(\Xibm\to\pim\Lb$) \cite{Aaij:2015yoy},
and using charmed-baryon lifetimes available at that time,
they predict
$\BF\left(\decXicz\right)=(0.19 \pm 0.07)\%$ for constructive interference
and $\BF\left(\decXicz\right)\lessapprox 0.01\%$ for destructive interference
between the SUUD  and WS contributions.
We have redone their calculation using updated lifetime measurements \cite{Aaij:2019lwg,PDG},
finding
$\BF\left(\decXicz\right)=(0.14 \pm 0.07)\%$ for constructive interference
and $\BF\left(\decXicz\right)\lessapprox (0.018\pm 0.015)\%$ for destructive interference.
Faller and Mannel, on the other hand, predict $\BF\left(\decXicz\right)<0.3\%$, an upper limit obtained by assuming
constructive interference \cite{Faller:2015oma}.
Finally, Cheng \etal predict $\BF\left(\decXicz\right) \sim 0.0087\%$,
assuming negative interference \cite{Cheng:2015ckx,*Cheng:1992ff}.
We have not updated these last predictions;
the effect would be to lower Faller and Mannel's positive interference prediction and raise the Cheng \etal negative one, giving somewhat better agreement with
Gronau and Rosner's predictions.

In this Letter we measure $\BF\left(\decXicz\right)$ using data collected by the LHCb detector, corresponding to  $3.8~\rm fb^{-1}$ of integrated luminosity in 13~\!TeV center-of-mass energy $pp$  collisions taken in 2017 and 2018. Natural units are used in this Letter with $c=\hbar=1$.  The LHCb detector is a
single-arm forward spectrometer covering the pseudorapidity range
$2 < \eta < 5$,
described in detail in Refs.~\cite{LHCb-DP-2008-001,LHCb-DP-2014-002}.
The trigger~\cite{LHCb-DP-2012-004}  consists of a
hardware stage, based on information from the calorimeter and muon
systems, followed by a software stage, which reconstructs charged particles.

Simulation is required to model the effects of the detector acceptance and selection requirements.
  We generate $pp$ collisions using
\pythia~\cite{Sjostrand:2007gs,*Sjostrand:2006za}
 with a specific \lhcb configuration~\cite{LHCb-PROC-2010-056}.  Decays of unstable particles
are described by \evtgen~\cite{Lange:2001uf}, where final-state
radiation is generated using \photos~\cite{Golonka:2005pn}. The
interaction of the particles with the detector, and its response,
are implemented using the \geant
toolkit~\cite{Allison:2006ve, *Agostinelli:2002hh} as described in
Ref.~\cite{LHCb-PROC-2011-006}.

In our analysis we use the prompt $\Xicz$ sample i.e. baryons, and their excitations, produced directly in the $pp$ collisions. Measurement of $\BF(\decXicz)$ is hampered by the lack of accurately measured $\Xicz$ branching fractions \cite{PDG} to be used for normalization. A measurement of $\BF(\PXi_c^0\to \pi^+\PXi^-)$ with a 29\% uncertainty exists \cite{Li:2018qak}, but the efficiency for reconstructing  $\PXi^-$ baryons is low in LHCb, in particular without a dedicated trigger line, so using this mode would lead to an unacceptably large error.
We overcome this difficulty by using two indirect methods, described below, that require additional measurements of prompt $\Lc$ and $\Xicp$ yields, both reconstructed in the $pK^-\pi^+$ decay mode. The same decay mode is also used to reconstruct $\Lc$ from the \decXicz decays.

We use a two-step process to maximize the statistical significance of our signal channel, as well as the two normalization channels. First, we apply a set of loose selection criteria to obtain samples with large signal efficiencies and suppressed background.
Subsequently, we use three different boosted decision trees~(BDT)~\cite{Breiman,AdaBoost}, one for each baryon decay, implemented in the TMVA toolkit~\cite{Hocker:2007ht,*Stelzer:2008zz}, to further separate signal from background.

The loose selection criteria for the $pK^-\pi^+$ final states include requirements on the tracks to have sufficient transverse momenta (\pt), be separated from the primary $pp$ collision vertex (PV), form a three-track vertex, and be identified as the hypothesized particle species. For the $\decXicz$ decay we require, in addition, that the $pK^-\pi^+$ has a mass within $\pm$20\mev of the \Lc mass peak; that there is an additional $\pi^-$ meson, which when combined with the \Lc candidate, has an invariant mass  from $-85$\mev below the known $\Xicz$ mass \cite{PDG} to 115\mev above; and that the \pt of the $\Xicz$ candidate is greater than 5\gev.

The BDTs are trained with background samples from data and simulated signal samples. Background training samples for the $\Lc$ and $\Xicp$ candidates are taken from the sideband regions on both sides of the mass peaks. For the $\Lc$ baryon background the intervals are  $40-65\mev$ away from the known $\Lc$ mass \cite{PDG}. For the $\Xicp$ baryon training the lower and higher sidebands are taken $40-58\mev$ and $40-72\mev$ from the known $\Xicp$ mass \cite{PDG}, respectively.
The $\Xicz$ background is constructed from like-sign $\pi^+\Lc$ candidates within $\pm 5\mev$ of the known $\Xicz$ baryon mass \cite{PDG}. For the $\Lc$ and $\Xicp$ candidates, we compute the $pK^-\pi^+$ invariant mass after constraining the three decay particles to form a common vertex and the summed momentum vector to point to the PV; this fitter is referred to as the ``Decay Tree Fitter" (DTF) \cite{Hulsbergen:2005pu}.  In the case of the $\Xicz$ baryon we add the additional $\pi^{\mp}$~meson before performing the fit. Only 1/10 of the available $\Lc\to p K^-\pi^+$ data sample is used to measure the $\Lc$ yield due to the large samples available relative to the other channels.

The variables used in the $\Lc$ and $\Xicp$ BDTs are the particle identification probabilities; the
$\chi^2_{\rm IP}$ of the $pK^-\pi^+$  with respect to the primary vertex, where
 $\chi^2_{\rm IP}$ is defined as the difference in the vertex fit $\chi^2$ with and without the $p$, $K^-$, and $\pi^+$ tracks; the angle between the particle's momentum vector and the vector from the original PV before the DTF refitting to the particle's decay vertex; the decay distance from the PV, and the DTF $\chisq$. The $\Xicz$ candidates are selected by a separate BDT using the same criteria used for the \Lc by adding similar extra variables associated with the additional pion.

\begin{figure}[b]
\vspace{-2mm}
  \begin{center}
\includegraphics[width=0.5\linewidth]{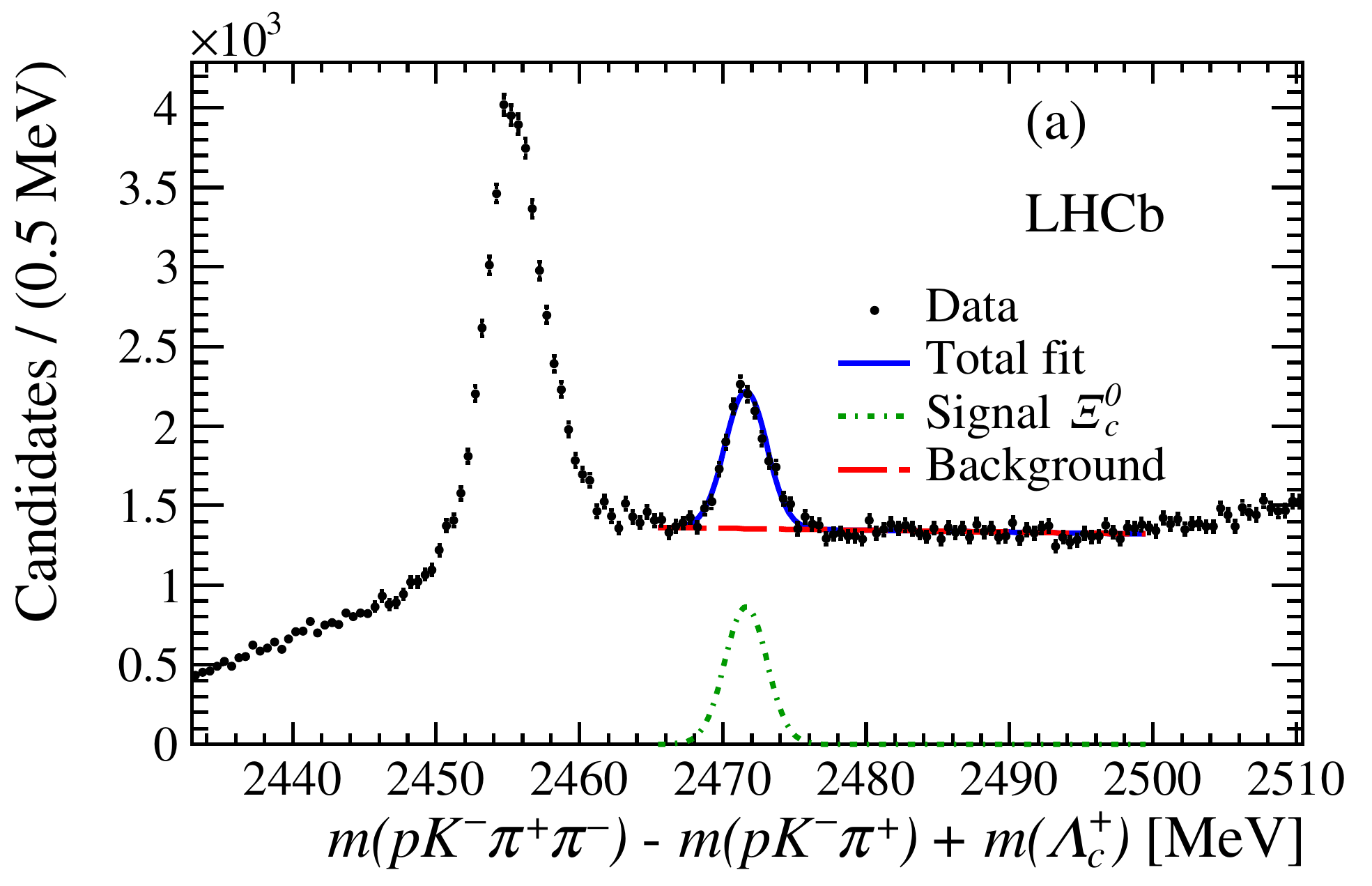}\includegraphics[width=0.5\linewidth]{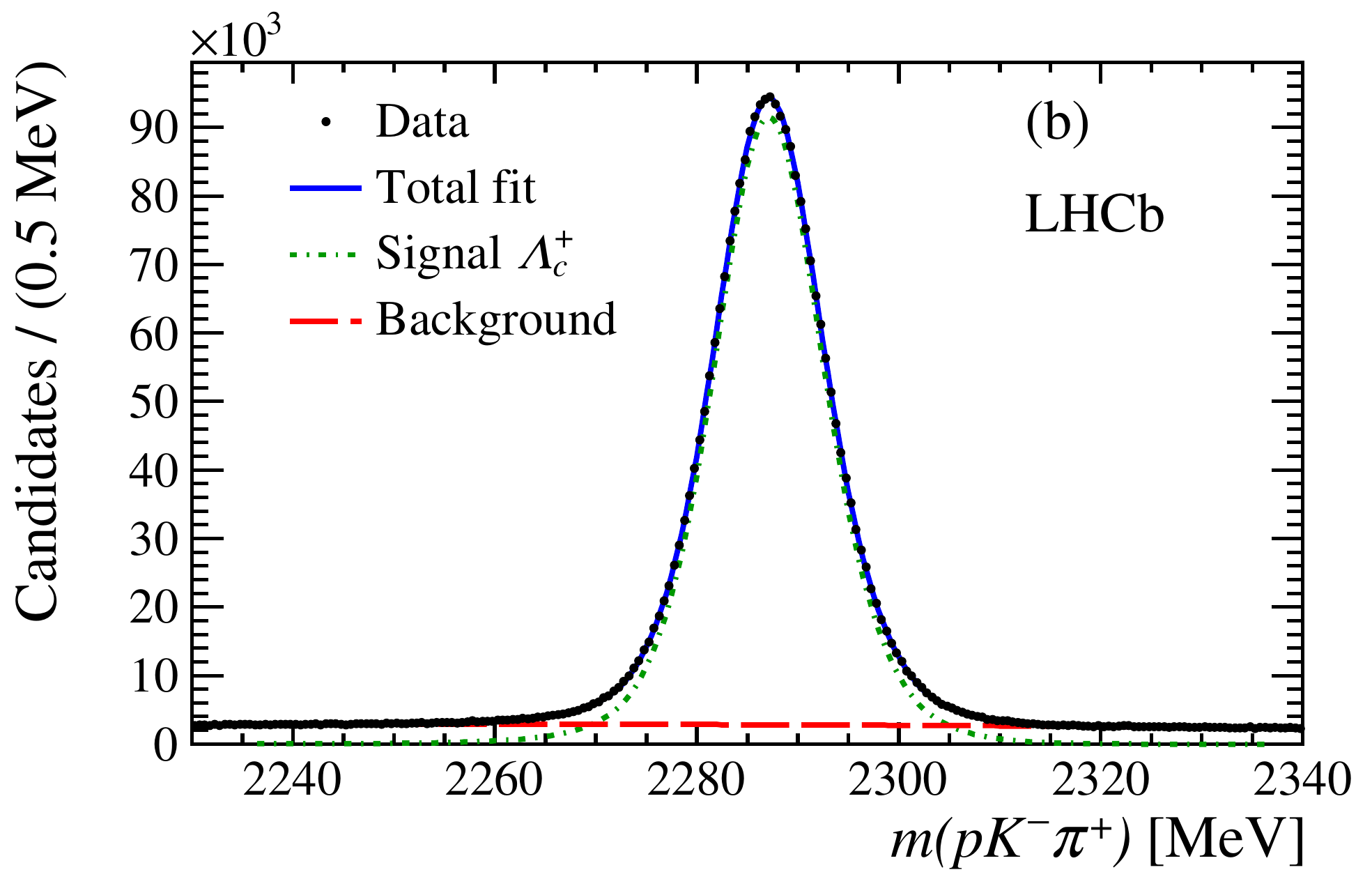}
\includegraphics[width=0.5\linewidth]{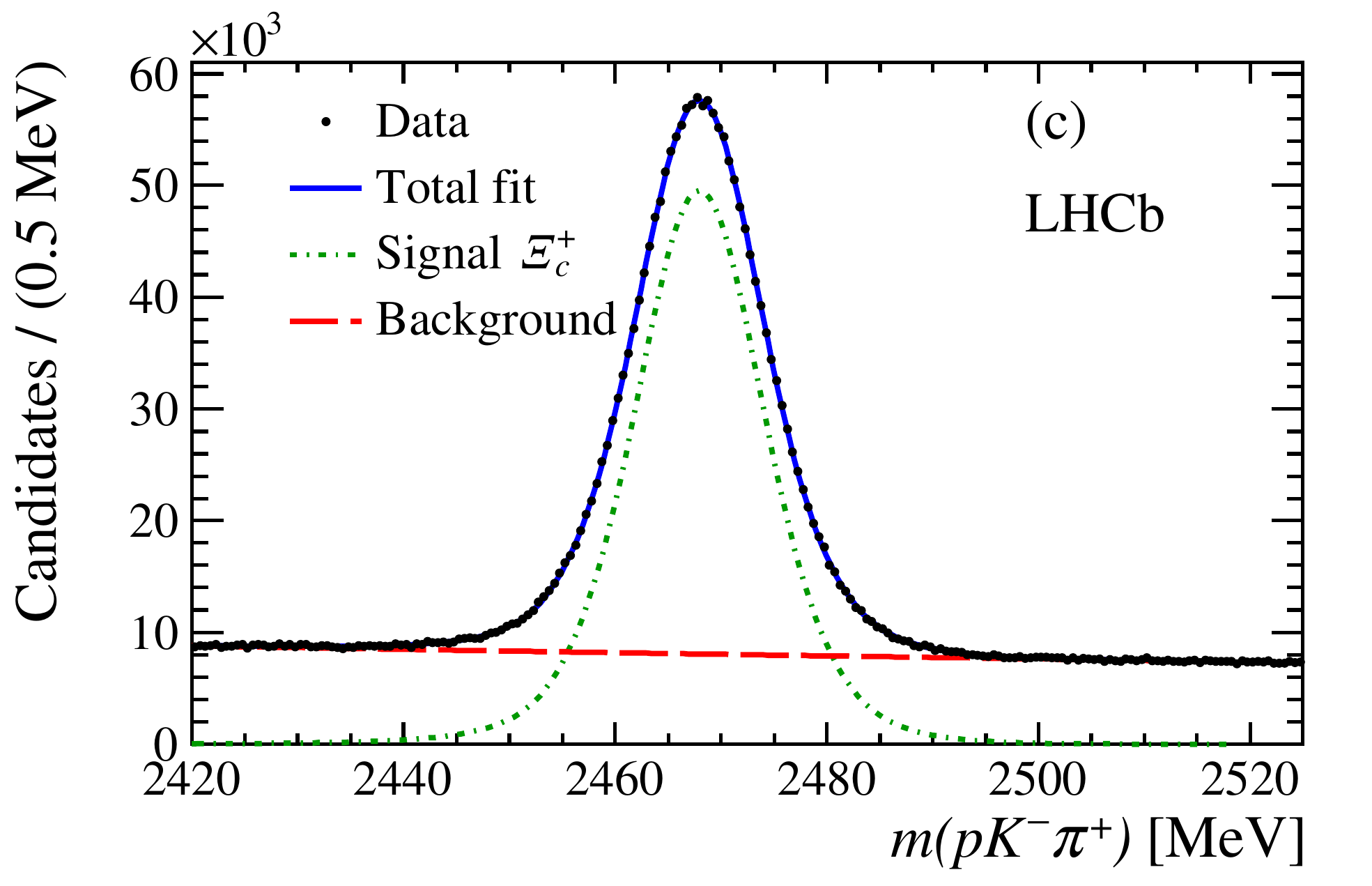}
  \end{center}
  \caption{Reconstructed invariant-mass distributions and signal fits of (a) $m(pK^-\pi^+\pi^-)$  showing a large $\PSigma_c^0$ signal with a smaller $\PXi_c^0$ signal, (b) $m(pK^-\pi^+)$ showing the $\Lc$ signal, and (c) $m(pK^-\pi^+)$ showing the $\PXi_c^+$ signal. For (a) the signal shape is a Crystal Ball function \cite{Skwarnicki:1986xj} with a high-mass tail, and the background shape is linear. For (b) and (c) the signal shapes are double-sided Crystal Ball plus single Gaussian functions, while the background shapes are second-order polynomials. The data in (b) only use 1/10 of the available sample.}
   \label{dataMass}
\end{figure}

The BDT selections are optimized by maximizing the ratio of signal efficiency to the square root of the number of candidates in the regions where we expect signal peaks. We show the resulting mass spectra in Fig.~\ref{dataMass}; the data are fitted using the signal and background shapes described in the figure caption. The fit yields are $6\!~320\pm230$ $\Xicz$, $2\!~667\!~200\pm3\!~300$ $\Lc$, and $1\!~613\!~000\pm3\!~500$ $\Xicp$ signal decays. To take into account the efficiency variation we perform the fits in four bins, two in \pt and two in $\eta$, and apply efficiencies calculated in each bin.

Trigger efficiencies are estimated from data, using the technique described in Ref.~\cite{Tolk:1701134}. Selection efficiencies are determined using simulated events, which are weighted to reproduce the resonance structures in the $pK^-\pi^+$ final states visible in the $\Lc$ and $\Xicp$ signal samples. The overall detection efficiencies are $(0.11\pm 0.02)$\%, $\left[(0.35\pm0.01)/10\right]$\%, and $(1.18\pm 0.03)\%$ for $\Xicz,~\Lc,$ and $\Xicp$ decays, respectively, where the factor of 10 is the prescale.

The first normalization method uses the \lhcb measurement of the relative production fractions of the $\Xibm$ and $\Lb$ beauty baryons,
${f_{\Xibm}}/{f_{\Lb}}=(8.2\pm 0.7\pm 2.6)\%$ \cite{Aaij:2019ezy}. Using HQS we equate the unmeasured production ratio of $\Xicz$ to $\Lc$ baryons,
$f_{\Xicz}/f_{\Lc}$,  to $\mathcal{C}\cdot f_{\Xibm}/f_{\Lb}$, where $\mathcal{C}$ is a correction factor for feed-downs of excited $\Xib$ baryons that do not have equal rates to $\Xibm$ and $\Xibz$ final states.
This feed-down is not symmetric primarily because the $\PXi_b^{\prime}(5935)^0$ state always decays to $\pi^0~(\rm{or}~\gamma)$~$\PXi_b^0$ \cite{Aaij:2016jnn}, since its mass is too low to decay into $\pi^+\PXi_b^-$. On the other hand, both the \Xibprm and \Xibstm states are seen to decay into both $\pi^-\Xibz$ and $\pi^0\Xibm$ final states  \cite{Aaij:2014yka}.  Any not yet observed higher mass states would be isospin symmetric in their decays.
Accounting for all the known excited states, and the associated phase-space corrections,  results in $\mathcal{C}= 1.18\pm 0.04$, where the uncertainty arises from the errors on the relative branching fraction measurements.

The second method uses the recent Belle measurement $\BF(\decXicp)=(0.45 \pm 0.21 \pm 0.07)\%$ \cite{Li:2019atu}.
Here we take the production of $\Xicz$ baryons equal to that of $\Xicp$ by isospin symmetry, e.g. $f_{\Xicz}/f_{\Xicp}=1.00\pm0.01$ \cite{PDG-Iso}. As the final state particles in the $\Xicp$ decay are the same  as in the $\Lc$ decay, many systematic uncertainties cancel.

We determine $ \BF(\decXicz)$ using the two measured ratios
\begin{align}
             \mathcal{R}_1 \equiv  \frac{N(\Xicz)}{N(\Lc)} &=\frac{f_\Xicz}{f_\Lc} \cdot \BF(\decXicz)
                               =
                              (0.095 \pm 0.003 \pm 0.012)\%,\nonumber \\
            \mathcal{R}_2 \equiv   \frac{N(\Xicz)}{N(\Xicp)} & =\frac{f_\Xicz}{f_\Xicp} \cdot \frac{\BF(\decLc)}{\BF(\decXicp)} \cdot \BF(\decXicz)
                             =
                            (5.70 \pm 0.19 \pm 0.77)\%,  \nonumber
\end{align}
where $N(i)$ indicates the efficiency corrected number of signal events for baryon $i$, $f_i$ indicates the fraction of particle production with respect to all $c$- or $b$-quark production, and the uncertainties are statistical and systematic, respectively, a convention used in the rest of this Letter. As discussed above, ${f_\Xicz}/{f_\Lc}=\mathcal{C}\cdot f_{\Xibm}/f_{\Lb}=(9.7\pm 0.9\pm 3.1)$\%, where we have added a 5\% relative systematic uncertainty, explained later, to account for our assumption of HQS.

We also determine $\BF(\decXicp)$ using
\begin{equation}
\mathcal{R}_3 \equiv \frac{N(\Xicp)}{N(\Lc)}
=  \frac{f_\Xicp}{f_\Lc} \cdot \frac{\BF(\decXicp)}{\BF(\decLc)} = (1.753 \pm 0.003 \pm 0.107)\%, \nonumber
\end{equation}
where $\BF(\decLc)=(6.23 \pm 0.33)\%$ \cite{PDG}.
The correlation matrix for these three results is
\begin{equation}
  \label{eq:intrinsic_correlation}
  \begin{pmatrix}
                  & \mathcal{R}_1 & \mathcal{R}_2 & \mathcal{R}_3 \\
    \mathcal{R}_1 & 1             & 0.71         & ~~0.15         \\
    \mathcal{R}_2 & ...         & 1             & -0.18        \\
    \mathcal{R}_3 &...        & ...        & ~1             \\
  \end{pmatrix}\nonumber
\end{equation}
The derived branching fractions are
\begin{align}
  \mathcal{B}_1 &\equiv \BF(\decXicz) = (0.98 \pm 0.04 \pm 0.35)\%,\nonumber\\
  \mathcal{B}_2 &\equiv \BF(\decXicz) = (0.41 \pm 0.01 \pm 0.21)\%,\nonumber\\
  \mathcal{B}_3 &\equiv \BF(\decXicp) = (1.135 \pm 0.002 \pm 0.387)\%.\nonumber
\end{align}
Their correlation matrix is
\begin{equation}
  \label{eq:derived_correlation}
  \begin{pmatrix}
                  & \mathcal{B}_1 & \mathcal{B}_2 & \mathcal{B}_3 \\
    \mathcal{B}_1 & 1             & 0.07         & ~~0.92         \\
    \mathcal{B}_2 & ...         & 1             & -0.02        \\
    \mathcal{B}_3 & ...         &...        & ~1             \\
  \end{pmatrix}\nonumber
\end{equation}
The weighted average value of  $\mathcal{B}_1$ and $\mathcal{B}_2$, taking into account their correlated error, is
\begin{equation}
  \label{eq:resultXic0_mean}
 {\BF(\decXicz)} = (0.55\pm 0.02 \pm 0.18)\%.\nonumber
\end{equation}

Systematic uncertainties dominate these results due to our reliance on external inputs. Our assumption of HQS to relate $f_{\Xicz}/f_{\Lc}$ to $f_{\Xibm}/f_{\Lb}$ is justified by considering the  analogous ratios of production fractions between charm and beauty states in 13\!~TeV \proton\proton collisions,
%\begin{equation}
%\label{eq:frac}
$\frac{f_{\Ds}}{f_{\Dz}+f_{\Dp}}$ and $\frac{f_{\Bs}}{f_{\Bz}+f_{\Bp}}$.
%\end{equation}
The beauty ratio is measured using semimuonic decays into a charmed meson, determined in the kinematic range $4<\pt<25\gev$\!, and is equal to $0.122\pm 0.006$ \cite{Aaij:2019pqz}.
Using the total charm cross-sections reported for $0<\pt<15\gev$ in Ref.~\cite{LHCb-PAPER-2015-041}, we find
%\begin{equation}
$\frac{f_{\Ds}}{f_{\Dz}+f_{\Dp}}\approx 0.121,$
%\end{equation}
where the statistical uncertainty is negligible.
The  systematic uncertainties in the charm-meson ratio including tracking, particle identification, luminosity, \etc,  mostly cancel. The uncertainties in the charm meson branching fractions cancel in the comparison with the $B$ meson ratio, because the same values are used in both. Thus we are left with a few percent uncertainty in the comparison of the charm and beauty meson ratios. The \pt distributions of the ratios are somewhat different; they fall linearly in the beauty case~\cite{Aaij:2019pqz} and are flatter in the charm case~\cite{LHCb-PAPER-2015-041}.
Taking this into account, a 5\% relative uncertainty due to the HQS assumption appears reasonable. Contamination of the charm baryons from $b$-decay sources is estimated in simulation and subtracted. The resultant systematic uncertainties in the ratios are small.
Table~\ref{tab:syserr} summarizes the sources of systematic uncertainty.

\begin{table}[b]
\centering
\caption{Systematic uncertainties in the branching fraction measurements. Ghost tracks refers to uncertainties from falsely reconstructed tracks. PID refers to particle identification efficiencies. Intermediate decays refers to the uncertainties  caused by inexact modeling of the resonant structures in the charmed-baryon decays. The $b$-decay sources refer to charmed baryons originating
from $b$-baryon decays included in our primarily prompt samples.  Relative $\int\lum$ refers to minor differences in the accumulated luminosities of the data samples for each of the three decays. The summed uncertainties are obtained by adding the individual components in quadrature.}
\vspace{0.2cm}
\begin{tabular}{lccc}
\hline\hline
Source & \multicolumn{3}{c}{Estimate (\%)}                     \\
       & \multicolumn{2}{c}{$\BF(\decXicz)$} & $\BF(\decXicp)$ \\
       & $\mathcal{B}_1$ & $\mathcal{B}_2$                  &  $\mathcal{B}_3$                \\
\hline
$f_\Xibm / f_\Lb$                                     & 32~~ & -- & 32~~ \\
$f_\Xicz / f_\Lc = \mathcal{C} \cdot f_\Xibm / f_\Lb$ & 6  & -- & 6  \\
$f_\Xicz / f_\Xicp = 1$                               & -- & 1  & 1  \\
$\BF(\decXicp)$                                       & -- & 49~~ & -- \\
$\BF(\decLc)$                                         & -- & 5  & 5  \\
Simulation statistics                                 & 4  & 3  & 2  \\
Trigger efficiency                                     & 7 & 8 & 2  \\
Ghost tracks                                      & 2  & 2  & 0  \\
PID                                                   & 1  & 1  & 1  \\
Tracking efficiencies                                            & 2  & 2  & 0  \\
Fit yields                                             & 6  & 6  & 3  \\
Intermediate decays                                         & 2  & 2  & 2  \\
$b$-decay sources                                          & 2  & 0  & 2  \\
Lifetimes                                             & 3  & 3  & 2  \\
Relative $\int\lum$                                   & -- & 1  & 1  \\
\hline
Sum of external                                       & 33~~ & 49~~ & 33~~ \\
\hline
Sum of intrinsic                                      & 12~~ & 13~~ & 6  \\
\hline
Sum of all                                            & 35~~ & 51~~ & 34~~ \\
\hline\hline
\end{tabular}
\label{tab:syserr}.
\end{table}

\begin{figure}[t]
  \begin{center}
    \includegraphics[width=0.8\linewidth]{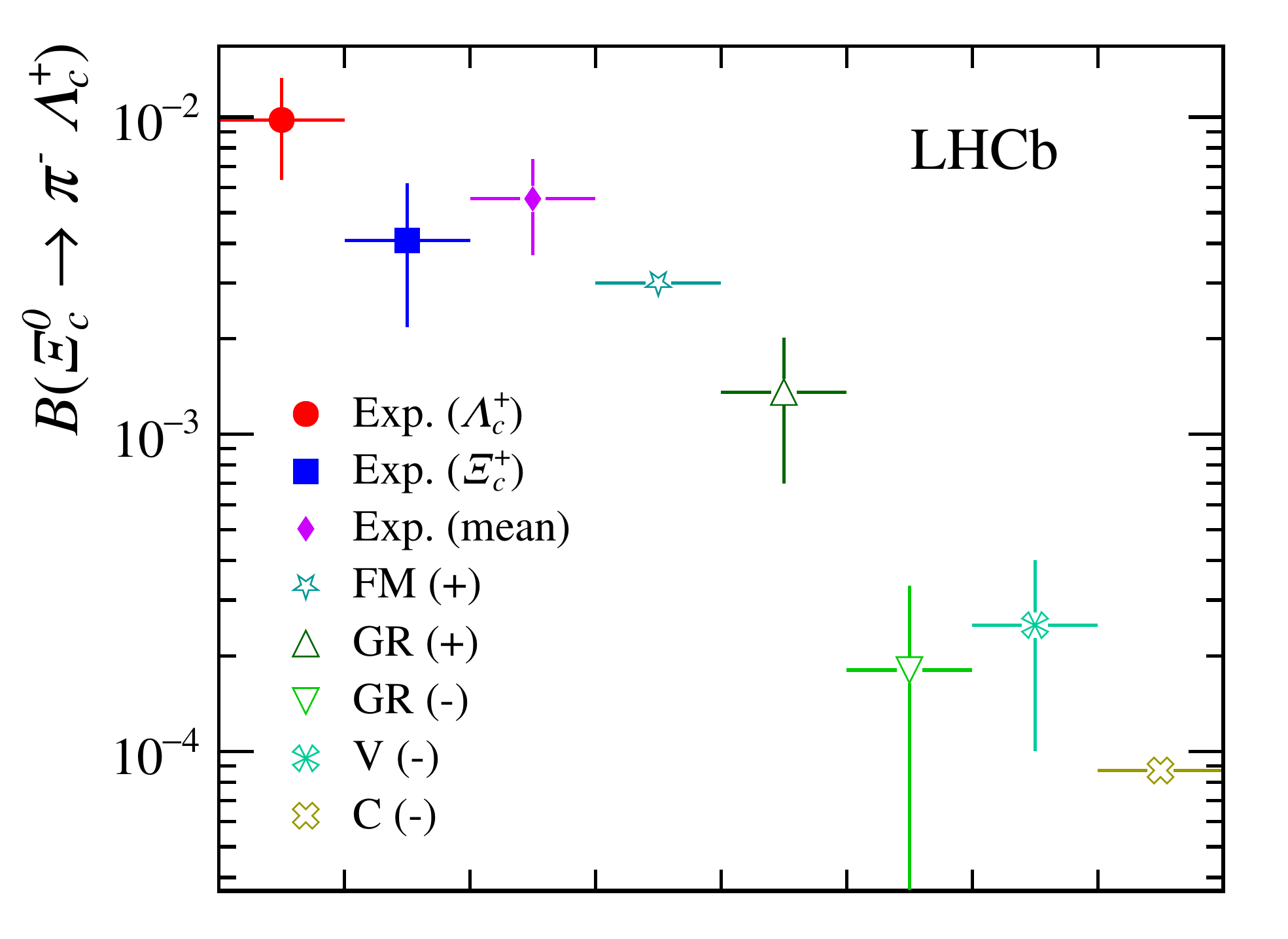}
  \end{center}
  \caption{Comparison of our two measurements of $\BF\left(\decXicz\right)$, and their average,  with the lower limit of Voloshin (V)\cite{Voloshin:2019ngb}, the upper limit of Faller and Mannel \cite{Faller:2015oma} (FM),  updated predictions of Gronau and Rosner \cite{Gronau:2016xiq} (GR),  and Cheng \etal \cite{Cheng:2015ckx} (C). The (+ or $-$) indicates if positive or negative interference between the SUUD and WS amplitudes is assumed.
  }
  \label{compareGR}
\end{figure}

In conclusion,  we perform the first measurement of the branching fraction of the suppressed $\decXicz$ decays, giving ${\BF(\decXicz)} = (0.55\pm 0.02 \pm 0.18)\%$. We compare with the theoretical predictions in Fig.~\ref{compareGR}; while our measurements are somewhat larger, we are in agreement with Gronau and Rosner's constructive interference prediction. Our result is also consistent with the Faller and Mannel  upper limit arrived at by assuming constructive interference \cite{Faller:2015oma}. We disagree, however, with
Cheng's  prediction of $\BF\left(\decXicz\right)$ assuming negative interference \cite{Cheng:2015ckx}. In addition, the branching fraction of the normalization channel is found to be $\BF(\decXicp) = (1.135 \pm 0.002 \pm 0.387)$\%, that is somewhat larger than, but in agreement with a previous Belle measurement~\cite{Li:2019atu}, and has a better relative precision.

{\clearpage}

\section*{Acknowledgements} 
% These Acknowledgements valid from 3-May-2019
%
\noindent We express our gratitude to our colleagues in the CERN
accelerator departments for the excellent performance of the LHC. We
thank the technical and administrative staff at the LHCb
institutes.
We acknowledge support from CERN and from the national agencies:
CAPES, CNPq, FAPERJ and FINEP (Brazil); 
MOST and NSFC (China); 
CNRS/IN2P3 (France); 
BMBF, DFG and MPG (Germany); 
INFN (Italy); 
NWO (Netherlands); 
MNiSW and NCN (Poland); 
MEN/IFA (Romania); 
MSHE (Russia); 
MinECo (Spain); 
SNSF and SER (Switzerland); 
NASU (Ukraine); 
STFC (United Kingdom); 
DOE NP and NSF (USA).
We acknowledge the computing resources that are provided by CERN, IN2P3
(France), KIT and DESY (Germany), INFN (Italy), SURF (Netherlands),
PIC (Spain), GridPP (United Kingdom), RRCKI and Yandex
LLC (Russia), CSCS (Switzerland), IFIN-HH (Romania), CBPF (Brazil),
PL-GRID (Poland) and OSC (USA).
We are indebted to the communities behind the multiple open-source
software packages on which we depend.
Individual groups or members have received support from
AvH Foundation (Germany);
EPLANET, Marie Sk\l{}odowska-Curie Actions and ERC (European Union);
A*MIDEX, ANR, Labex P2IO and OCEVU, and R\'{e}gion Auvergne-Rh\^{o}ne-Alpes (France);
Key Research Program of Frontier Sciences of CAS, CAS PIFI, and the Thousand Talents Program (China);
RFBR, RSF and Yandex LLC (Russia);
GVA, XuntaGal and GENCAT (Spain);
the Royal Society
and the Leverhulme Trust (United Kingdom).

\clearpage
\newpage
%\appendix
%\input{supplemental}

\addcontentsline{toc}{section}{References}
%%\setboolean{inbibliography}{true}
%\bibliographystyle{LHCb}
%\bibliography{Xic0-prl,LHCb-DP,main,LHCb-PAPER}

\begin{mcitethebibliography}{10}
\mciteSetBstSublistMode{n}
\mciteSetBstMaxWidthForm{subitem}{\alph{mcitesubitemcount})}
\mciteSetBstSublistLabelBeginEnd{\mcitemaxwidthsubitemform\space}
{\relax}{\relax}

\bibitem{Voloshin:2000et}
M.~B. Voloshin, \ifthenelse{\boolean{articletitles}}{\emph{{Weak decays $\PXi
  (Q) \to \PLambda (Q) \pi$}},
  }{}\href{https://doi.org/10.1016/S0370-2693(00)00150-7}{Phys.\ Lett.\
  \textbf{B476} (2000) 297},
  \href{http://arxiv.org/abs/hep-ph/0001057}{{\normalfont\ttfamily
  arXiv:hep-ph/0001057}}\relax
\mciteBstWouldAddEndPuncttrue
\mciteSetBstMidEndSepPunct{\mcitedefaultmidpunct}
{\mcitedefaultendpunct}{\mcitedefaultseppunct}\relax
\EndOfBibitem
\bibitem{Treiman:1967cua}
S.~Treiman, \ifthenelse{\boolean{articletitles}}{\emph{{Commutator algebra and
  the PCAC hypothesis}}, }{}Comments Nucl.\ Part.\ Phys.\  \textbf{1} (1967)
  13\relax
\mciteBstWouldAddEndPuncttrue
\mciteSetBstMidEndSepPunct{\mcitedefaultmidpunct}
{\mcitedefaultendpunct}{\mcitedefaultseppunct}\relax
\EndOfBibitem
\bibitem{Aaij:2019lwg}
LHCb collaboration, R.~Aaij {\em et~al.},
  \ifthenelse{\boolean{articletitles}}{\emph{{Precision measurement of the
  $\Lc$, $\PXi_c^+$ and $\PXi_c^0$ baryon lifetimes}},
  }{}\href{https://doi.org/10.1103/PhysRevD.100.032001}{Phys.\ Rev.\
  \textbf{D100} (2019) 032001},
  \href{http://arxiv.org/abs/1906.08350}{{\normalfont\ttfamily
  arXiv:1906.08350}}\relax
\mciteBstWouldAddEndPuncttrue
\mciteSetBstMidEndSepPunct{\mcitedefaultmidpunct}
{\mcitedefaultendpunct}{\mcitedefaultseppunct}\relax
\EndOfBibitem
\bibitem{Aaij:2015yoy}
LHCb collaboration, R.~Aaij {\em et~al.},
  \ifthenelse{\boolean{articletitles}}{\emph{{Evidence for the
  strangeness-changing weak decay $\PXi_b^-\to\Lb\pi^-$}},
  }{}\href{https://doi.org/10.1103/PhysRevLett.115.241801}{Phys.\ Rev.\ Lett.\
  \textbf{115} (2015) 241801},
  \href{http://arxiv.org/abs/1510.03829}{{\normalfont\ttfamily
  arXiv:1510.03829}}\relax
\mciteBstWouldAddEndPuncttrue
\mciteSetBstMidEndSepPunct{\mcitedefaultmidpunct}
{\mcitedefaultendpunct}{\mcitedefaultseppunct}\relax
\EndOfBibitem
\bibitem{Voloshin:2019ngb}
M.~B. Voloshin, \ifthenelse{\boolean{articletitles}}{\emph{{Update on splitting
  of lifetimes of $c$ and $b$ hyperons within the heavy quark expansion and
  decays $\PXi_Q \to \PLambda_Q \pi$}},
  }{}\href{https://doi.org/10.1103/PhysRevD.100.114030}{Phys.\ Rev.\ D
  \textbf{100} (2019) 114030},
  \href{http://arxiv.org/abs/1911.05730}{{\normalfont\ttfamily
  arXiv:1911.05730}}\relax
\mciteBstWouldAddEndPuncttrue
\mciteSetBstMidEndSepPunct{\mcitedefaultmidpunct}
{\mcitedefaultendpunct}{\mcitedefaultseppunct}\relax
\EndOfBibitem
\bibitem{Gronau:2016xiq}
M.~Gronau and J.~L. Rosner, \ifthenelse{\boolean{articletitles}}{\emph{{From
  $\PXi_b \to \Lb \pi$ to $\PXi_c \to \Lc \pi$}},
  }{}\href{https://doi.org/10.1016/j.physletb.2016.04.020}{Phys.\ Lett.\
  \textbf{B757} (2016) 330},
  \href{http://arxiv.org/abs/1603.07309}{{\normalfont\ttfamily
  arXiv:1603.07309}}\relax
\mciteBstWouldAddEndPuncttrue
\mciteSetBstMidEndSepPunct{\mcitedefaultmidpunct}
{\mcitedefaultendpunct}{\mcitedefaultseppunct}\relax
\EndOfBibitem
\bibitem{PDG}
Particle Data Group, M.~Tanabashi {\em et~al.},
  \ifthenelse{\boolean{articletitles}}{\emph{{\href{http://pdg.lbl.gov/}{Review
  of particle physics}}},
  }{}\href{https://doi.org/10.1103/PhysRevD.98.030001}{Phys.\ Rev.\
  \textbf{D98} (2018) 030001}, and {\href{http://pdglive.lbl.gov/}{2019
  update}}\relax
\mciteBstWouldAddEndPuncttrue
\mciteSetBstMidEndSepPunct{\mcitedefaultmidpunct}
{\mcitedefaultendpunct}{\mcitedefaultseppunct}\relax
\EndOfBibitem
\bibitem{Faller:2015oma}
S.~Faller and T.~Mannel,
  \ifthenelse{\boolean{articletitles}}{\emph{{Light-quark decays in heavy
  hadrons}}, }{}\href{https://doi.org/10.1016/j.physletb.2015.09.072}{Phys.\
  Lett.\  \textbf{B750} (2015) 653},
  \href{http://arxiv.org/abs/1503.06088}{{\normalfont\ttfamily
  arXiv:1503.06088}}\relax
\mciteBstWouldAddEndPuncttrue
\mciteSetBstMidEndSepPunct{\mcitedefaultmidpunct}
{\mcitedefaultendpunct}{\mcitedefaultseppunct}\relax
\EndOfBibitem
\bibitem{Cheng:2015ckx}
H.-Y. Cheng {\em et~al.},
  \ifthenelse{\boolean{articletitles}}{\emph{{Heavy-flavor-conserving hadronic
  weak wecays of heavy baryons}},
  }{}\href{https://doi.org/10.1007/JHEP03(2016)028}{JHEP \textbf{03} (2016)
  028}, \href{http://arxiv.org/abs/1512.01276}{{\normalfont\ttfamily
  arXiv:1512.01276}}\relax
\mciteBstWouldAddEndPuncttrue
\mciteSetBstMidEndSepPunct{\mcitedefaultmidpunct}
{\mcitedefaultendpunct}{\mcitedefaultseppunct}\relax
\EndOfBibitem
\bibitem{Cheng:1992ff}
H.-Y. Cheng {\em et~al.}, \ifthenelse{\boolean{articletitles}}{\emph{{Heavy
  flavor conserving nonleptonic weak decays of heavy baryons}},
  }{}\href{https://doi.org/10.1103/PhysRevD.46.5060}{Phys.\ Rev.\ D \textbf{46}
  (1992) 5060}\relax
\mciteBstWouldAddEndPuncttrue
\mciteSetBstMidEndSepPunct{\mcitedefaultmidpunct}
{\mcitedefaultendpunct}{\mcitedefaultseppunct}\relax
\EndOfBibitem
\bibitem{LHCb-DP-2008-001}
LHCb collaboration, A.~A. Alves~Jr.\ {\em et~al.},
  \ifthenelse{\boolean{articletitles}}{\emph{{The \lhcb detector at the LHC}},
  }{}\href{https://doi.org/10.1088/1748-0221/3/08/S08005}{JINST \textbf{3}
  (2008) S08005}\relax
\mciteBstWouldAddEndPuncttrue
\mciteSetBstMidEndSepPunct{\mcitedefaultmidpunct}
{\mcitedefaultendpunct}{\mcitedefaultseppunct}\relax
\EndOfBibitem
\bibitem{LHCb-DP-2014-002}
LHCb collaboration, R.~Aaij {\em et~al.},
  \ifthenelse{\boolean{articletitles}}{\emph{{LHCb detector performance}},
  }{}\href{https://doi.org/10.1142/S0217751X15300227}{Int.\ J.\ Mod.\ Phys.\
  \textbf{A30} (2015) 1530022},
  \href{http://arxiv.org/abs/1412.6352}{{\normalfont\ttfamily
  arXiv:1412.6352}}\relax
\mciteBstWouldAddEndPuncttrue
\mciteSetBstMidEndSepPunct{\mcitedefaultmidpunct}
{\mcitedefaultendpunct}{\mcitedefaultseppunct}\relax
\EndOfBibitem
\bibitem{LHCb-DP-2012-004}
R.~Aaij {\em et~al.}, \ifthenelse{\boolean{articletitles}}{\emph{{The \lhcb
  trigger and its performance in 2011}},
  }{}\href{https://doi.org/10.1088/1748-0221/8/04/P04022}{JINST \textbf{8}
  (2013) P04022}, \href{http://arxiv.org/abs/1211.3055}{{\normalfont\ttfamily
  arXiv:1211.3055}}\relax
\mciteBstWouldAddEndPuncttrue
\mciteSetBstMidEndSepPunct{\mcitedefaultmidpunct}
{\mcitedefaultendpunct}{\mcitedefaultseppunct}\relax
\EndOfBibitem
\bibitem{Sjostrand:2007gs}
T.~Sj\"{o}strand, S.~Mrenna, and P.~Skands,
  \ifthenelse{\boolean{articletitles}}{\emph{{A brief introduction to PYTHIA
  8.1}}, }{}\href{https://doi.org/10.1016/j.cpc.2008.01.036}{Comput.\ Phys.\
  Commun.\  \textbf{178} (2008) 852},
  \href{http://arxiv.org/abs/0710.3820}{{\normalfont\ttfamily
  arXiv:0710.3820}}\relax
\mciteBstWouldAddEndPuncttrue
\mciteSetBstMidEndSepPunct{\mcitedefaultmidpunct}
{\mcitedefaultendpunct}{\mcitedefaultseppunct}\relax
\EndOfBibitem
\bibitem{Sjostrand:2006za}
T.~Sj\"{o}strand, S.~Mrenna, and P.~Skands,
  \ifthenelse{\boolean{articletitles}}{\emph{{PYTHIA 6.4 physics and manual}},
  }{}\href{https://doi.org/10.1088/1126-6708/2006/05/026}{JHEP \textbf{05}
  (2006) 026}, \href{http://arxiv.org/abs/hep-ph/0603175}{{\normalfont\ttfamily
  arXiv:hep-ph/0603175}}\relax
\mciteBstWouldAddEndPuncttrue
\mciteSetBstMidEndSepPunct{\mcitedefaultmidpunct}
{\mcitedefaultendpunct}{\mcitedefaultseppunct}\relax
\EndOfBibitem
\bibitem{LHCb-PROC-2010-056}
I.~Belyaev {\em et~al.}, \ifthenelse{\boolean{articletitles}}{\emph{{Handling
  of the generation of primary events in Gauss, the LHCb simulation
  framework}}, }{}\href{https://doi.org/10.1088/1742-6596/331/3/032047}{{J.\
  Phys.\ Conf.\ Ser.\ } \textbf{331} (2011) 032047}\relax
\mciteBstWouldAddEndPuncttrue
\mciteSetBstMidEndSepPunct{\mcitedefaultmidpunct}
{\mcitedefaultendpunct}{\mcitedefaultseppunct}\relax
\EndOfBibitem
\bibitem{Lange:2001uf}
D.~J. Lange, \ifthenelse{\boolean{articletitles}}{\emph{{The EvtGen particle
  decay simulation package}},
  }{}\href{https://doi.org/10.1016/S0168-9002(01)00089-4}{Nucl.\ Instrum.\
  Meth.\  \textbf{A462} (2001) 152}\relax
\mciteBstWouldAddEndPuncttrue
\mciteSetBstMidEndSepPunct{\mcitedefaultmidpunct}
{\mcitedefaultendpunct}{\mcitedefaultseppunct}\relax
\EndOfBibitem
\bibitem{Golonka:2005pn}
P.~Golonka and Z.~Was, \ifthenelse{\boolean{articletitles}}{\emph{{PHOTOS Monte
  Carlo: A precision tool for QED corrections in $Z$ and $W$ decays}},
  }{}\href{https://doi.org/10.1140/epjc/s2005-02396-4}{Eur.\ Phys.\ J.\
  \textbf{C45} (2006) 97},
  \href{http://arxiv.org/abs/hep-ph/0506026}{{\normalfont\ttfamily
  arXiv:hep-ph/0506026}}\relax
\mciteBstWouldAddEndPuncttrue
\mciteSetBstMidEndSepPunct{\mcitedefaultmidpunct}
{\mcitedefaultendpunct}{\mcitedefaultseppunct}\relax
\EndOfBibitem
\bibitem{Allison:2006ve}
Geant4 collaboration, J.~Allison {\em et~al.},
  \ifthenelse{\boolean{articletitles}}{\emph{{Geant4 developments and
  applications}}, }{}\href{https://doi.org/10.1109/TNS.2006.869826}{IEEE
  Trans.\ Nucl.\ Sci.\  \textbf{53} (2006) 270}\relax
\mciteBstWouldAddEndPuncttrue
\mciteSetBstMidEndSepPunct{\mcitedefaultmidpunct}
{\mcitedefaultendpunct}{\mcitedefaultseppunct}\relax
\EndOfBibitem
\bibitem{Agostinelli:2002hh}
Geant4 collaboration, S.~Agostinelli {\em et~al.},
  \ifthenelse{\boolean{articletitles}}{\emph{{Geant4: A simulation toolkit}},
  }{}\href{https://doi.org/10.1016/S0168-9002(03)01368-8}{Nucl.\ Instrum.\
  Meth.\  \textbf{A506} (2003) 250}\relax
\mciteBstWouldAddEndPuncttrue
\mciteSetBstMidEndSepPunct{\mcitedefaultmidpunct}
{\mcitedefaultendpunct}{\mcitedefaultseppunct}\relax
\EndOfBibitem
\bibitem{LHCb-PROC-2011-006}
M.~Clemencic {\em et~al.}, \ifthenelse{\boolean{articletitles}}{\emph{{The
  \lhcb simulation application, Gauss: Design, evolution and experience}},
  }{}\href{https://doi.org/10.1088/1742-6596/331/3/032023}{{J.\ Phys.\ Conf.\
  Ser.\ } \textbf{331} (2011) 032023}\relax
\mciteBstWouldAddEndPuncttrue
\mciteSetBstMidEndSepPunct{\mcitedefaultmidpunct}
{\mcitedefaultendpunct}{\mcitedefaultseppunct}\relax
\EndOfBibitem
\bibitem{Li:2018qak}
Belle collaboration, Y.~B. Li {\em et~al.},
  \ifthenelse{\boolean{articletitles}}{\emph{{First measurements of absolute
  branching fractions of the $\PXi_c^0$ baryon at Belle}},
  }{}\href{https://doi.org/10.1103/PhysRevLett.122.082001}{Phys.\ Rev.\ Lett.\
  \textbf{122} (2019) 082001},
  \href{http://arxiv.org/abs/1811.09738}{{\normalfont\ttfamily
  arXiv:1811.09738}}\relax
\mciteBstWouldAddEndPuncttrue
\mciteSetBstMidEndSepPunct{\mcitedefaultmidpunct}
{\mcitedefaultendpunct}{\mcitedefaultseppunct}\relax
\EndOfBibitem
\bibitem{Breiman}
L.~Breiman, J.~H. Friedman, R.~A. Olshen, and C.~J. Stone, {\em Classification
  and regression trees}, Wadsworth international group, Belmont, California,
  USA, 1984\relax
\mciteBstWouldAddEndPuncttrue
\mciteSetBstMidEndSepPunct{\mcitedefaultmidpunct}
{\mcitedefaultendpunct}{\mcitedefaultseppunct}\relax
\EndOfBibitem
\bibitem{AdaBoost}
R.~E. Schapire and Y.~Freund, \ifthenelse{\boolean{articletitles}}{\emph{A
  decision-theoretic generalization of on-line learning and an application to
  boosting}, }{}\href{https://doi.org/10.1006/jcss.1997.1504}{Journal Computer
  and System Sciences \textbf{55} (1997) 119}\relax
\mciteBstWouldAddEndPuncttrue
\mciteSetBstMidEndSepPunct{\mcitedefaultmidpunct}
{\mcitedefaultendpunct}{\mcitedefaultseppunct}\relax
\EndOfBibitem
\bibitem{Hocker:2007ht}
A.~Hoecker {\em et~al.}, \ifthenelse{\boolean{articletitles}}{\emph{{TMVA:
  Toolkit for multivariate data analysis}}, }{}PoS \textbf{ACAT} (2007) 040,
  \href{http://arxiv.org/abs/physics/0703039}{{\normalfont\ttfamily
  arXiv:physics/0703039}}\relax
\mciteBstWouldAddEndPuncttrue
\mciteSetBstMidEndSepPunct{\mcitedefaultmidpunct}
{\mcitedefaultendpunct}{\mcitedefaultseppunct}\relax
\EndOfBibitem
\bibitem{Stelzer:2008zz}
J.~Stelzer, A.~Hocker, P.~Speckmayer, and H.~Voss,
  \ifthenelse{\boolean{articletitles}}{\emph{{Current developments in TMVA: An
  outlook to TMVA4}}, }{}\href{https://doi.org/10.22323/1.070.0063}{PoS
  \textbf{ACAT08} (2008) 063}\relax
\mciteBstWouldAddEndPuncttrue
\mciteSetBstMidEndSepPunct{\mcitedefaultmidpunct}
{\mcitedefaultendpunct}{\mcitedefaultseppunct}\relax
\EndOfBibitem
\bibitem{Hulsbergen:2005pu}
W.~D. Hulsbergen, \ifthenelse{\boolean{articletitles}}{\emph{{Decay chain
  fitting with a Kalman filter}},
  }{}\href{https://doi.org/10.1016/j.nima.2005.06.078}{Nucl.\ Instrum.\ Meth.\
  \textbf{A552} (2005) 566},
  \href{http://arxiv.org/abs/physics/0503191}{{\normalfont\ttfamily
  arXiv:physics/0503191}}\relax
\mciteBstWouldAddEndPuncttrue
\mciteSetBstMidEndSepPunct{\mcitedefaultmidpunct}
{\mcitedefaultendpunct}{\mcitedefaultseppunct}\relax
\EndOfBibitem
\bibitem{Skwarnicki:1986xj}
T.~Skwarnicki, {\em {A study of the radiative cascade transitions between the
  Upsilon-prime and Upsilon resonances}}, PhD thesis, Institute of Nuclear
  Physics, Krakow, 1986,
  {\href{http://inspirehep.net/record/230779/}{DESY-F31-86-02}}\relax
\mciteBstWouldAddEndPuncttrue
\mciteSetBstMidEndSepPunct{\mcitedefaultmidpunct}
{\mcitedefaultendpunct}{\mcitedefaultseppunct}\relax
\EndOfBibitem
\bibitem{Tolk:1701134}
S.~Tolk, J.~Albrecht, F.~Dettori, and A.~Pellegrino,
  \ifthenelse{\boolean{articletitles}}{\emph{{Data driven trigger efficiency
  determination at LHCb}}, }{}
  {\href{https://cds.cern.ch/record/1701134}{LHCb-PUB-2014-039,
  CERN-LHCb-PUB-2014-039}}, May, 2014\relax
\mciteBstWouldAddEndPuncttrue
\mciteSetBstMidEndSepPunct{\mcitedefaultmidpunct}
{\mcitedefaultendpunct}{\mcitedefaultseppunct}\relax
\EndOfBibitem
\bibitem{Aaij:2019ezy}
LHCb collaboration, R.~Aaij {\em et~al.},
  \ifthenelse{\boolean{articletitles}}{\emph{{Measurement of the mass and
  production rate of $\PXi_b^-$ baryons}},
  }{}\href{https://doi.org/10.1103/PhysRevD.99.052006}{Phys.\ Rev.\
  \textbf{D99} (2019) 052006},
  \href{http://arxiv.org/abs/1901.07075}{{\normalfont\ttfamily
  arXiv:1901.07075}}\relax
\mciteBstWouldAddEndPuncttrue
\mciteSetBstMidEndSepPunct{\mcitedefaultmidpunct}
{\mcitedefaultendpunct}{\mcitedefaultseppunct}\relax
\EndOfBibitem
\bibitem{Aaij:2016jnn}
LHCb collaboration, R.~Aaij {\em et~al.},
  \ifthenelse{\boolean{articletitles}}{\emph{{Measurement of the properties of
  the $\Xires_b^{\ast 0}$ baryon}},
  }{}\href{https://doi.org/10.1007/JHEP05(2016)161}{JHEP \textbf{05} (2016)
  161}, \href{http://arxiv.org/abs/1604.03896}{{\normalfont\ttfamily
  arXiv:1604.03896}}\relax
\mciteBstWouldAddEndPuncttrue
\mciteSetBstMidEndSepPunct{\mcitedefaultmidpunct}
{\mcitedefaultendpunct}{\mcitedefaultseppunct}\relax
\EndOfBibitem
\bibitem{Aaij:2014yka}
LHCb collaboration, R.~Aaij {\em et~al.},
  \ifthenelse{\boolean{articletitles}}{\emph{{Observation of two new $\Xibm$
  baryon resonances}},
  }{}\href{https://doi.org/10.1103/PhysRevLett.114.062004}{Phys.\ Rev.\ Lett.\
  \textbf{114} (2015) 062004},
  \href{http://arxiv.org/abs/1411.4849}{{\normalfont\ttfamily
  arXiv:1411.4849}}\relax
\mciteBstWouldAddEndPuncttrue
\mciteSetBstMidEndSepPunct{\mcitedefaultmidpunct}
{\mcitedefaultendpunct}{\mcitedefaultseppunct}\relax
\EndOfBibitem
\bibitem{Li:2019atu}
Belle collaboration, Y.~B. Li {\em et~al.},
  \ifthenelse{\boolean{articletitles}}{\emph{{First measurements of absolute
  branching fractions of the $\PXi_c^+$ baryon at Belle}},
  }{}\href{https://doi.org/10.1103/PhysRevD.100.031101}{Phys.\ Rev.\
  \textbf{D100} (2019) 031101},
  \href{http://arxiv.org/abs/1904.12093}{{\normalfont\ttfamily
  arXiv:1904.12093}}\relax
\mciteBstWouldAddEndPuncttrue
\mciteSetBstMidEndSepPunct{\mcitedefaultmidpunct}
{\mcitedefaultendpunct}{\mcitedefaultseppunct}\relax
\EndOfBibitem
\bibitem{PDG-Iso}
Particle Data Group, M.~Tanabashi {\em et~al.},
  \ifthenelse{\boolean{articletitles}}{\emph{{\href{http://pdg.lbl.gov/}{Review
  of particle physics}}},
  }{}\href{https://doi.org/10.1103/PhysRevD.98.030001}{Phys.\ Rev.\
  \textbf{D98} (2018) 030001}, and {\href{http://pdglive.lbl.gov/}{2019 update;
  see section 84.2}}\relax
\mciteBstWouldAddEndPuncttrue
\mciteSetBstMidEndSepPunct{\mcitedefaultmidpunct}
{\mcitedefaultendpunct}{\mcitedefaultseppunct}\relax
\EndOfBibitem
\bibitem{Aaij:2019pqz}
LHCb collaboration, R.~Aaij {\em et~al.},
  \ifthenelse{\boolean{articletitles}}{\emph{{Measurement of $b$-hadron
  fractions in 13 TeV $pp$ collisions}},
  }{}\href{http://arxiv.org/abs/1902.06794}{{\normalfont\ttfamily
  arXiv:1902.06794}}\relax
\mciteBstWouldAddEndPuncttrue
\mciteSetBstMidEndSepPunct{\mcitedefaultmidpunct}
{\mcitedefaultendpunct}{\mcitedefaultseppunct}\relax
\EndOfBibitem
\bibitem{LHCb-PAPER-2015-041}
LHCb collaboration, R.~Aaij {\em et~al.},
  \ifthenelse{\boolean{articletitles}}{\emph{{Measurements of prompt charm
  production cross-sections in $\proton\proton$ collisions at $\sqs =
  $13\tev}}, }{}\href{https://doi.org/10.1007/JHEP03(2016)159}{JHEP \textbf{03}
  (2016) 159}, Erratum \href{https://doi.org/10.1007/JHEP09(2016)013}{ibid.\
  \textbf{09} (2016) 013}, Erratum
  \href{https://doi.org/10.1007/JHEP05(2017)074}{ibid.\   \textbf{05} (2017)
  074}, \href{http://arxiv.org/abs/1510.01707}{{\normalfont\ttfamily
  arXiv:1510.01707}}\relax
\mciteBstWouldAddEndPuncttrue
\mciteSetBstMidEndSepPunct{\mcitedefaultmidpunct}
{\mcitedefaultendpunct}{\mcitedefaultseppunct}\relax
\EndOfBibitem
\end{mcitethebibliography}
\ifx\mcitethebibliography\mciteundefinedmacro
\PackageError{LHCb.bst}{mciteplus.sty has not been loaded}
{This bibstyle requires the use of the mciteplus package.}\fi
\providecommand{\href}[2]{#2}

\newpage
% LHCb collaboration author list
% Data extracted on July 21st, 2020 at 3:14pm for reference date 12-Jun-2020
\centerline
{\large\bf LHCb collaboration}
\begin
{flushleft}
\small
R.~Aaij$^{31}$,
C.~Abell{\'a}n~Beteta$^{49}$,
T.~Ackernley$^{59}$,
B.~Adeva$^{45}$,
M.~Adinolfi$^{53}$,
H.~Afsharnia$^{9}$,
C.A.~Aidala$^{84}$,
S.~Aiola$^{25}$,
Z.~Ajaltouni$^{9}$,
S.~Akar$^{64}$
% ...

{\footnotesize \it

% ...
$ ^{9}$Universit{\'e} Clermont Auvergne, CNRS/IN2P3, LPC, Clermont-Ferrand, France\\
% ...
$ ^{25}$INFN Sezione di Milano, Milano, Italy\\
% ...
$ ^{31}$Nikhef National Institute for Subatomic Physics, Amsterdam, Netherlands\\
% ...
$ ^{45}$Instituto Galego de F{\'\i}sica de Altas Enerx{\'\i}as (IGFAE), Universidade de Santiago de Compostela, Santiago de Compostela, Spain\\
% ...
$ ^{49}$Physik-Institut, Universit{\"a}t Z{\"u}rich, Z{\"u}rich, Switzerland\\
% ...
$ ^{53}$H.H. Wills Physics Laboratory, University of Bristol, Bristol, United Kingdom\\
% ...
$ ^{59}$Oliver Lodge Laboratory, University of Liverpool, Liverpool, United Kingdom\\
% ...
$ ^{64}$University of Cincinnati, Cincinnati, OH, United States\\
% ...
$ ^{67}$Syracuse University, Syracuse, NY, United States\\
% ...
$ ^{84}$University of Michigan, Ann Arbor, United States, associated to $^{67}$\\
% ...
\medskip
}
\end{flushleft}

\end{document}